\documentclass[slac_one]{revtex4}
\usepackage{graphicx}
\usepackage{fancyhdr}
\begin{document}

\title{Precision charmonium and D physics from lattice QCD and determination 
of the charm quark mass} %% Paper title goes here


\author{C. T. H. Davies (for the HPQCD collaboration)}
\affiliation{Department of Physics and Astronomy, University of Glasgow, Glasgow G12 8QQ, UK}
%
%\author{P. Lucas}
%\affiliation{FNAL, Batavia, IL 60510, USA}

\begin{abstract}
I will describe recent results from the HPQCD collaboration using a 
new very accurate method for charm quarks in lattice QCD, 
that we have used in calculations including the full effect 
of $u$, $d$ and $s$ sea quarks. Multiple values of the lattice spacing and of the 
$u$, $d$ and $s$ sea quark masses allow us to extrapolate reliably, with a 
full error budget, to the real world. 
This opens up the field of charm physics to precision lattice QCD tests. 
So far we have calculated the $D$ and $D_s$ meson masses to 6 MeV, 
having fixed the charm quark mass from the $\eta_c$ meson. 
Our $D$ and $D_s$ decay constants (determined to 2\%) 
make an interesting comparison to CLEO-c results as we await 
improved experimental errors. 
We are also able to determine the charm quark mass to an accuracy of 1\% 
using charmonium correlators and high-order continuum QCD perturbation theory. 
Future calculations are briefly discussed. 
\end{abstract}

\maketitle

\thispagestyle{fancy}


\section{INTRODUCTION} % Section title should be in all capitals.
Lattice QCD is now established as a precision tool for 
`gold-plated' hadron physics~\cite{orig}, enabling
us both to test QCD to high accuracy but also to use 
it to determine parameters of the Standard Model
such as quark masses and CKM matrix 
elements. 

\begin{figure*}[t]
\centering
\includegraphics[width=105mm,angle=-90]{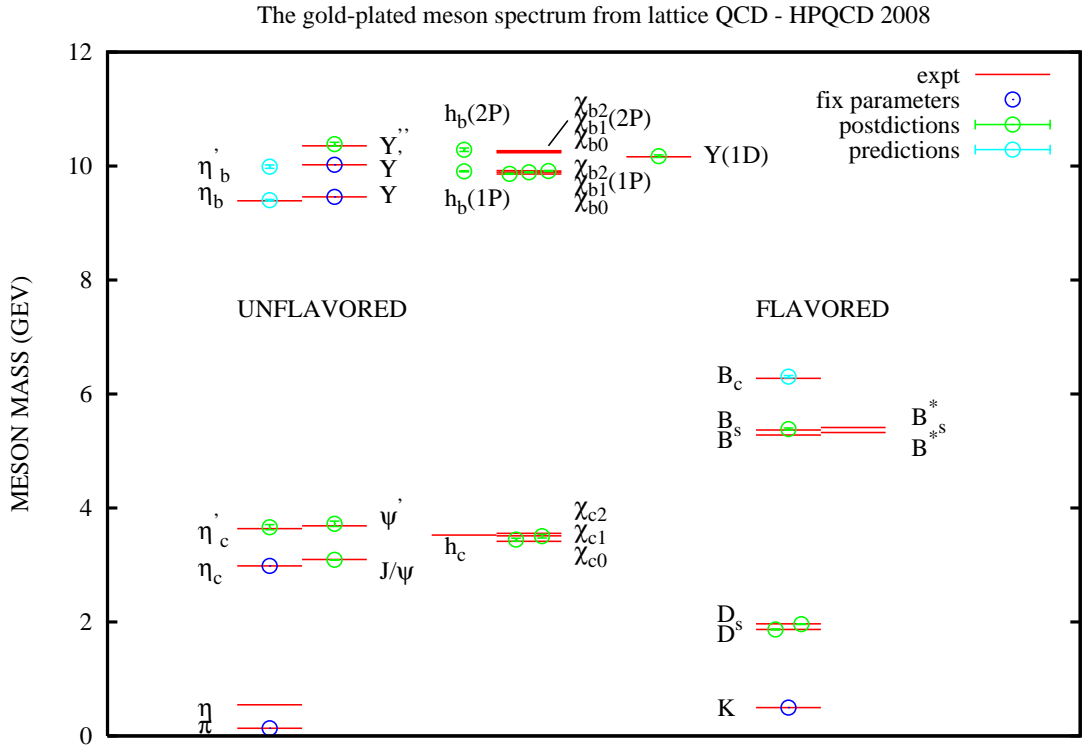}
\caption{The gold-plated meson spectrum from lattice QCD. Lines represent 
experimental values. Open circles are lattice results, distinguishing 
those hadron masses used to fix the quark mass parameters and those 
which were predictions ahead of experiment~\cite{ups,bc}. } \label{fig:goldspect}
\end{figure*}

Figure~\ref{fig:goldspect} shows the current status 
of HPQCD lattice calculations of the masses of gold-plated
mesons. These are mesons which are far from strong 
decay thresholds and whose masses can then be accurately 
determined both theoretically and experimentally.  
The lattice calculations are done including the full 
effect of sea $u$, $d$ and $s$ quarks (using the 
MILC collaboration gluon configurations) and at multiple 
values of the lattice spacing so that systematic errors 
from working on a space-time lattice can be removed 
as far as possible, and a full error budget worked out. 
A lot of the gold-plated mesons contain valence 
$c$ or $b$ quarks. Whilst work continues on 
$\Upsilon$, $B$ and related mesons there has been a lot 
of recent progress on mesons containing 
$c$ quarks, and I report on that here. 
One of the key issues in charm physics is to 
test lattice QCD against experiment as a precursor
to using accurate lattice QCD results for $B$ physics
along with experiment to determine key CKM elements.
We now have a 2\% accurate lattice QCD calculation 
of the decay constants of the $D$ and $D_s$ mesons~\cite{fdshort} 
and the recent experimental updates on this 
from CLEO-c have thrown up an interesting picture.
This is discussed in section 2. 
Further theoretical tests of our charm methods 
have led to a 1\% accurate determination of the 
charm quark mass, using methods pioneered in 
continnum QCD~\cite{kuhn}, and this is discussed in section 3. 

\section{CHARM QUARKS IN LATTICE QCD}

In continuum QCD both charm and bottom quarks are described 
as heavy because their masses are significantly greater 
than the QCD scale, $\Lambda_{QCD}$, of a few hundred MeV. 
Special techniques can be applied to heavy quarks because
they are nonrelativistic in their bound 
states and $\alpha_s(m_Q)$ is relatively small. 
This is important information that can be applied 
to handling $b$ quarks in lattice 
QCD also. For $c$ quarks it turns out not
to be nearly so useful because the charm quark mass 
in lattice units, $m_ca$, is not that large for values 
of the lattice spacing currently in use, and will get 
even smaller as we make finer and finer lattices. 
We believe that a better strategy for $c$ quarks is 
to treat them in the same way as light $u$, $d$ and $s$ quarks. 
This allows us to make use of light quark symmetries of 
the (lattice) QCD Lagrangian so that, for example, 
the annihilation rate of pseudoscalar mesons can 
be directly calculated without the need for 
a renormalisation constant. 
We can also apply standard variance reduction technqiues and 
find then that lattice statistical errors for $D/D_s$ 
mesons are as small (less than 1\%) as those for $\pi$ and $K$. 

The worry, however, with $c$ quarks, is that $m_ca$ is 
still quite large and this means that $m_ca$ will 
set the scale for discretisation errors 
that appear as a result of having a finite 
lattice spacing. Although these can be extrapolated 
away with multiple values of the lattice spacing, a
large extrapolation will lead to a large resulting error. 
Typically we would expect, for some quantity $m$, to 
obtain a result at non-zero lattice spacing, $a$, that 
behaves as: 
\begin{equation}\label{eq:disc}
m = m_{a=0}(1+A(m_ca)^2+B(m_ca)^4+\ldots ).
\end{equation}
For $a \approx$ 0.1fm, $(m_ca) \approx 0.4$ and $(m_ca)^2$ 
terms could lead to a 20\% error. These are present in 
most light quark actions (where they do not cause a big problem). 
We designed an action in which these 
terms were not present so that errors of only a few 
percent would be made even for $c$ quarks, and once extrapolations 
were made to zero lattice spacings any remaining errors 
would be very small. 
The action is based on the improved staggered quark action 
used for sea quarks in the MILC configurations, and 
is called the Highly Improved Staggered Quark action (HISQ)~\cite{hisq}. 

\subsection{Results for $D$ and $D_s$ mesons}

\begin{figure*}[t]
\centering
\includegraphics[width=85mm,angle=-90]{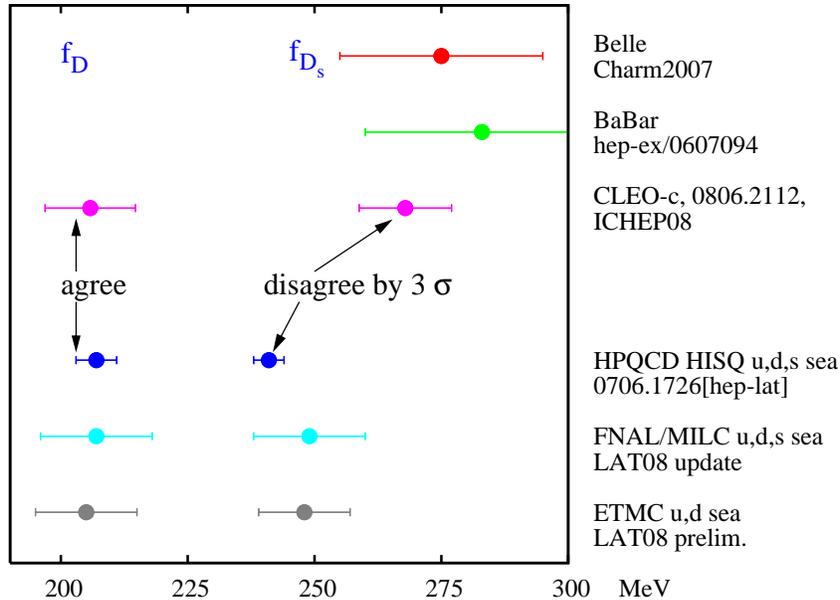}
\caption{A comparison of lattice results for the $D$ (leftmost points) 
and $D_s$ (rightmost points) decay
constants with experiment. The experimental results are obtained from the leptonic 
decay rate using CKM elements $V_{cs}$ and $V_{cd}$ from elsewhere~\cite{belle, babar, cleo-c}. 
CLEO-c presented an update at this conference~\cite{cleo-c-ichep}.
HPQCD results are from~\cite{fdshort} using HISQ quarks. 
The Fermilab/MILC results are an update this year~\cite{fnal-lat08}, ETMC 
results are new~\cite{etmc} and
do not include the $s$ sea quarks that are present in the real world. 
Neither of these results use a formalism as accurate 
as HISQ and hence have larger errors. 
Currently there is agreement between lattice and experiment for 
$f_D$ but not for $f_{D_s}$, when comparing HPQCD and CLEO-c results. } \label{fig:fds}
\end{figure*}

A lattice calculation proceeds by generating sets of gluon 
fields that are `typical snapshots of the vacuum'. 
The MILC collaboration now have many such sets of gluon 
fields at 5 different values of the lattice spacing 
including the full effect of $u$, $d$ and $s$ sea quarks 
using the improved staggered quark formalism~\cite{milc-lat07}. 
We have used $c$ and light valence HISQ quarks on 
these gluon configurations to calculate correlators 
for $\eta_c$, $D$, $D_s$, $K$ and $\pi$ mesons. 
By studying the correlators as a function of lattice 
timein the large time limit, we can then calculate the masses of these 
mesons and the decay amplitude known as the decay 
constant. This is defined in pure QCD as $f_H$ where
$f_Hm_H = <0|\overline{\psi}\gamma_0\gamma_5\psi|H>$,
and is the probability that the quark 
and antiquark are in the same place to annihilate 
to a $W$ boson. When calculated in QCD, it takes 
into account all the strong interactions that keep 
the quark and antiquark bound into the meson. 

The quark masses need to be fixed in any QCD calculation and 
we do this by adjusting them until the $\eta_c$, $K$ and $\pi$ 
meson masses are correct. The last two require
extrapolation in the $u/d$ quark mass (the sea and valence 
masses are chosen to be the same) using chiral perturbation 
theory because the lattice calculation has to be done at 
$u/d$ masses that are larger than the physical values. 
$D$ and $D_s$ masses have no further free parameters and 
we obtain values for them in agreement with experiment 
and with 6 MeV errors (again after extrapolation in 
the $u/d$ mass to the physical point)~\cite{fdshort}. This is a very strong 
test both of our control of discretisation errors and 
of QCD itself because it shows that charmonium and 
charm-light systems, with very different dynamics, 
are simultaneously described a single QCD Lagrangian.   

The rate for annihilation to a $W (\rightarrow l \overline{\nu}$)
is proportional to $f_H^2V_{ab}^2$ where $V_{ab}$ is the 
appropriate CKM element. Our results for $f_{K}/f_{\pi}$ 
mesons can be used, with experimental rates, to determine 
$V_{us}$ to 0.6\%~\cite{fdshort}. Our results for $D$ and $D_s$ 
decay constants have 2\% errors and can be compared to experimental 
determinations obtained by dividing the leptonic rate 
by known values of $V_{cs}$ and $V_{cd}$. 
Following our calculations the results from the 
experiment have become very exciting this year, as shown 
in Figure~\ref{fig:fds}. A new result for $f_D$ from 
CLEO-c~\cite{cleo-c} agrees well with our result, but 
values for $f_{D_s}$ do not. The discrepancy there amounts 
to 3$\sigma$ where $\sigma$ comes from experiment because 
our error is so small. This 
has led to speculation of new physics~\cite{kronfeld}, since 
the disagreement is a serious one and is the only 
such disagreement from the 15 or so quantities that 
have now been accurately calculated in lattice QCD and 
compared to experiment, see Figure~\ref{fig:goldspect}. 
Final results from CLEO-c on $f_{D_s}$ should reduce the 
error bars further and either confirm, or not, this 
hypothesis. 

Meanwhile, we have continued to test our control 
of lattice systematic errors by calculating 
vector (e.g. $J/\psi$) electromagnetic decay rates 
and hyperfine splittings such as $J/\psi - \eta_c$ and 
$D_s^*-D_s$. These are in good agreement with experiment and we are 
finalising error budgets for them. Future work aims at few percent 
errors for semileptonic form factors for $D \rightarrow K, \pi$ for 
comparison with experiment. 

\section{THE CHARM QUARK MASS}

We have developed, in collaboration with continuum QCD theorists, 
a new method for determining heavy quark masses. So far we have 
applied this to determine $m_c$ to 1\% and the agreement between 
our result for the mass in the $\overline{MS}$ scheme at 3 GeV, 
0.986(10) GeV ~\cite{mcjj}, and the result obtained by analogous continuum 
techniques, 0.986(13) GeV ~\cite{mccont}, is another strong test of our control of 
systematic errors in handling charm quarks in lattice QCD. 

Lattice QCD calculations encompass energy scales from low-energies and long-distances
(which we use to extract hadron masses) to relatively high-energies set by the 
inverse of the lattice spacing. These high-energies are well into the perturbative 
regime and we make use of this, for example, in the accurate determination 
of $\alpha_s$~\cite{alpha}. Our charmonium ($\eta_c$ or $J/\psi$) 
correlators are also perturbative at short lattice times, before the 
ground state meson dominates (and we extract the ground state meson mass). 
We can access this perturbative region by taking 'time-moments', 
$G_n = \sum_t (t/a)^n G(t)$, of the correlator, G(t). 
Extrapolation of these to the continuum allows them to be 
compared to continuum QCD perturbation theory which should 
be accurate for small enough values of $n$ and is known 
in some cases up to and including $\alpha_s^3$ terms. 
The analogous continuum calculation~\cite{mccont} effectively determines
correlator moments in the vector case from $R(e^+e^- \rightarrow hadrons)$ in 
the charm region. 
Our calculation in the vector ($J/\psi$) case agrees with experiment but 
our most accurate result, quoted above, comes from the pseudoscalar ($\eta_c$)~\cite{mcjj}. 

The determination of $m_c$ from the more conventional lattice method 
of converting the bare lattice mass to the $\overline{MS}$ scheme using 
lattice QCD perturbation theory, agrees well with this result~\cite{allison}.
The accurate results for $m_c$ can be leveraged into an accurate 
result for light quark masses by determining ratios such 
as $m_c/m_s$ from which renormalisation constants between the lattice 
and the continuum cancel~\cite{davieslepage}.

\begin{acknowledgments}
I am grateful to my colleagues in the HPQCD collaboration, and to 
members of the Fermilab lattice and MILC collaborations, for many 
useful discussions.
Work supported by the STFC and the 
Leverhulme Trust. 
\end{acknowledgments}

\end{document}